\documentclass[aoas,nameyear,dvips]{arximspdf}
\usepackage{dcolumn}
\usepackage{graphicx}

\doi{10.1214/10-AOAS424}
\volume{5}
\issue{2A}
\pubyear{2011}
\firstpage{746}
\lastpage{772}

\makeatletter
\newproclaim{asp}{Assumption}
\newcolumntype{d}[1]{D{.}{.}{#1}}

  \let\sv@tabnotetext\tabnotetext
  \let\sv@tabnotemark@fmt\tabnotemark@fmt
   \long\def\legend#1{{\let\tabnote@indent\leavevmode\sv@tabnotetext[]{}{#1}}}

\makeatother

\begin{document}
\begin{frontmatter}

\title{The effect of winning an Oscar Award on survival: Correcting for healthy performer survivor bias with a rank preserving
structural accelerated failure time model}

\runtitle{Survival in Oscar Award winning performers}

\begin{aug}
\author[A]{\fnms{Xu} \snm{Han}\corref{}\ead[label=e1]{xhan@princeton.edu}},
\author[B]{\fnms{Dylan S.} \snm{Small}\ead[label=e2]{dsmall@wharton.upenn.edu}},
\author[B]{\fnms{Dean P.} \snm{Foster}\ead[label=e3]{dean@foster.net}}
\and
\author[B]{\fnms{Vishal} \snm{Patel}\ead[label=e4]{vvp@wharton.upenn.edu}}

\runauthor{Han, Small, Foster and Patel}

\affiliation{Princeton University, University of Pennsylvania, University of
Pennsylvania and University of Pennsylvania}

\address[A]{X. Han\\
Department of Operations Research\\
\quad and Financial Engineering\\
Princeton University\\
Room 214, Sherrerd Hall\\
Princeton, New Jersey 08544\\USA\\
\printead{e1}} 

\address[B]{D. S. Small\\
D. P. Foster\\
V. Patel\\
Department of Statistics\\
Wharton School\\
University of Pennsylvania\\
400 Jon M. Huntsman Hall\\
3730 Walnut St.\\
Philadelphia, Pennsylvania 19104-6340\\
USA\\
\printead{e2}\\
\phantom{E-mail:\ }\printead*{e3}\\
\phantom{E-mail:\ }\printead*{e4}}
\end{aug}

\received{\smonth{8} \syear{2008}}
\revised{\smonth{9} \syear{2010}}

\begin{abstract}
We study the causal effect of winning an Oscar Award on an actor or
actress's survival. Does the increase in social rank from a performer
winning an Oscar increase the performer's life expectancy? Previous
studies of this issue have suffered from healthy performer survivor
bias, that is, candidates who are healthier will be able to act in more
films and have more chance to win Oscar Awards. To correct this bias,
we adapt Robins' rank preserving structural accelerated failure time
model and $g$-estimation method. We show in simulation studies that
this approach corrects the bias contained in previous studies. We
estimate that the effect of winning an Oscar Award on survival is 4.2
years, with a $95\%$ confidence interval of $[-0.4,8.4]$ years. There
is \textit{not} strong evidence that winning an Oscar increases life
expectancy.
\end{abstract}

\begin{keyword}
\kwd{Causal inference}
\kwd{survival analysis}
\kwd{Oscar Award}
\kwd{rank preserving structural accelerated failure time model}
\kwd{$g$-estimation.}
\end{keyword}

\end{frontmatter}

\section{Introduction}\label{s1}
Does an increase in a social animal's social ``rank'' cause the animal
to live longer? This question has been studied extensively in both
nonhuman primates and humans. Animals with social ranks that experience
more stress have been shown to experience adverse adrenocortical,
cardiovascular, reproductive, immunological, and neurobiological
consequences [Sapolsky (\citeyear{s2005})]. Redelmeier and Singh
(\citeyear{drs2001}) studied the impact of social rank on lifetime in an
intriguing context: among Hollywood actors and actresses, does winning
an Oscar Award (Academy Award) cause the actor's/actress's expected
lifetime to increase? In Redelmeier and Singh's most emphasized
comparison (the one cited in their abstract), they stated that life
expectancy was 3.9 years longer for Oscar Award winners than for other
less recognized performers and that this difference corresponded to a
$28\%$ mortality rate reduction for winners compared to less recognized
performers ($95\%$ CI: $10\%$ to $42\%$). In an interview, Dr.
Redelmeier stated, ``Once you've got that statuette on your mantel
place, it's an uncontested sign of peer approval that nobody can take
away from you, so that any subsequent harsh reviews leave you more
resilient. It doesn't quite get under your skin. The normal stresses
and strains of everyday life do not drag you down.'' [Associated Press
Story, February~26 (2005)].

In Redelmeier and Singh's analysis emphasized in their abstract, they
fit a Cox proportional hazards model with whether a performer ever wins
an Oscar Award in his or her lifetime treated as a time-independent
covariate and survival measured from the performer's date of birth.
Sylvestre, Huszti and Hanley (\citeyear{shh2006}) pointed out that
this analysis suffers from \textit{immortal time bias}---for a winner,
the time before winning is ``immortal time.'' In other words,
performers who live longer have more opportunities to win Oscar Awards.
To eliminate immortal time bias, Sylvestre et al. fit a Cox
proportional hazards model with winning status treated as a
time-dependent covariate and survival measured from a performer's date
of first nomination (Redelmeier and Singh also fit one time-dependent
covariate model with survival measured from the performer's date of
birth). Sylvestre et al. estimated that winning an Oscar Award had a
positive effect on lifetime, but the estimated effect was not
significant. Although a valuable step forward, Sylvestre et al.'s
analysis still suffers from \textit{healthy performer survivor bias}:
Candidates who are healthier will be able to act in more films and have
more chances to win Oscar Awards. We provide a more detailed
description of healthy performer survivor bias in Sections \ref{s2} and \ref{s3}.

In this paper we adapt James Robins' rank preserving structural
accelerated failure time model with $g$-estimation [Robins
(\citeyear{r1992}); Robins et al. (\citeyear{rbrw1992})] to eliminate
healthy performer survivor bias; it also eliminates immortal time bias,
which can be seen as one aspect of healthy performer survivor bias. Our
analysis is based on the assumption that the winner of each award is
selected randomly among the nominees conditional on age at time of
nomination, number of previous nominations and number of previous wins.
We first show in a simulation study the potential for healthy performer
survivor bias to make inferences from Cox models, with or without
time-dependent covariates, incorrect, and then show that $g$-estimation
provides correct inferences. We then analyze the effect of winning an
Oscar on life expectancy using $g$-estimation.

Our study also contributes to the debate that high socio-economic
status is associated with good health and long life. Famous examples
are the Whitehall studies of British civil servants; see Reid et al.
(\citeyear{rbhjkr1974}), Marmot, Rose and Hamilton
(\citeyear{mrh1978}), Marmot, Shipley and Rose (\citeyear{msr1984}),
Marmot et al. (\citeyear{mdspnhwbf1991}) and Ferrie et al.
(\citeyear{f2002}). Recently, Rablen and Oswald (\citeyear{ro2008})
studied the causal effect of winning a Nobel Prize on scientists'
longevity. Correcting for potential bias, they estimated that winning
the Nobel Prize, compared to merely being nominated, is associated with
between 1 and 2 years of extra longevity. Abel and Kruger
(\citeyear{ak2005}) studied the longevity of Baseball Hall of Famers
compared to the other players. They concluded that median
post-induction survival for Hall of Famers was 5 years shorter than for
noninducted players, which does not support the role of celebrity on
longevity.

The rest of our paper is organized as follows: Section \ref{s2} discusses
previous methods and their biases and presents a simulation study that
documents these biases, Section \ref{s3} describes the rank preserving
structural failure time model and $g$-estimation, Section \ref{s4} analyzes
the Oscar Award data and Section \ref{s5} provides conclusion and discussion.

\section{Existing methods and biases}\label{s2}
\subsection{Background for Oscar Awards}\label{s21}
The Oscar Awards are the most prominent and most watched film awards
ceremony in the world. They are presented annually by the Academy of
Motion Pictures Arts and Sciences. We will focus on the awards in four
categories---Best Lead Actor, Best Lead Actress, Best Supporting Actor,
and Best Supporting Actress. The annual awards selection process is
complex, but the brief schedule is as follows: In December, the Academy
compiles a list of eligible performers for an award. In January, all
Academy members nominate five performers in each of the four categories
(Best Lead Actor, Best Lead Actress, Best Supporting Actor, Best
Supporting Actress). In February, nominations for each performer are
tabulated, and the top five are publicly identified as nominees for
each category. Then all Academy members vote for one out of five
nominees, and the winner is the one who gets the most votes.

\subsection{Previous work}\label{s22}
Redelmeier and Singh (\citeyear{drs2001}) compiled a list of all
nominees for the Oscar Awards from 1929 to 2000 (72 years). They also
matched each nominee to a cast member who performed in the same film as
the nominee and was the same sex and born in the same era as the
nominee. Redelmeier and Singh's analysis was based on comparing 235
Oscar winners to 527 nonwinning nominees, and 887 performers who were
never nominated (controls). In their primary analysis, survival was
measured from performers' dates of birth.\footnote{Redelmeier and Singh
also considered survival from the day each performer's first film was
released, each performer's 65th birthday (excluding performers who died
before 65), and each performer's 50th birthday (excluding performers
who died before 50). As noted by Sylvestre, Huszti and Hanley
(\citeyear{shh2006}), all of these methods of measuring time-zero
suffer from immortal time bias.} In most of Redelmeier and Singh's
analyses, they used the winner status as a fixed-in-time covariate,
that is, a performer would be considered a winner throughout the study
if he or she won an Oscar Award at least once in his or her lifetime.
Kaplan--Meier curves showed that life expectancy was 3.9 years longer
for winner than for controls, and 3.5 years longer for winners than for
nonwinning nominees. In Cox proportional hazards models with no
adjustment for other covariates, winning was estimated to reduce
mortality by $28\%$ compared to controls and by $26\%$ compared to
nonwinning nominees, with lower $95\%$ confidence limits for both
comparisons greater than $0\%$, suggesting that winning an Oscar has a
beneficial effect on lifetime. Adjustment for demographic and
professional factors yielded similar results, with lower confidence
limits for the mortality reduction due to winning remaining above
$0\%$. Redelmeier and Singh considered one Cox proportional hazard
model that used the winner status as a time-dependent covariate, that
is, an Oscar Award winning performer is treated as a winner only after
he or she won an Award. This model estimated a mortality rate reduction
of $20\%$ for winners vs. controls, with a lower $95\%$ CI limit of
$0\%$.

Sylvestre, Huszti and Hanley (\citeyear{shh2006}) pointed out that
analyses that treat winner status as a fixed in time covariate credit
the winners' lifetime before winning toward survival subsequent to
winning. These ``immortal'' years will cause bias in the estimate of
the causal effect of winning. We will focus on Sylvestre et al.'s
method for correcting this bias in comparing winners to nonwinning
nominees. Sylvestre et al. used a Cox proportional hazard model that
differed in two ways from Redelmeier and Singh's primary analyses: (1)
winning was treated as a time-dependent covariate, an Oscar Award
winning performer only becomes a winner after he or she wins an award
(as noted above, Redelmeier and Singh also considered this approach in
one of their analyses); (2) a performer was only part of the risk set
once he or she was first nominated. Using this model, Sylvestre et al.
estimated a~mortality rate reduction of $18\%$ for winners vs.
nonwinning nominees with a $95\%$ CI of $-4\%$ to $35\%$. Thus, this
model estimates that winning an Oscar has a beneficial effect on
lifetime, but there is not strong evidence for a beneficial effect.
Note that Sylvestre et al. used an updated data set compared to
Redelmeier and Singh's; Sylvestre et al. considered a selection
interval for Oscar Awards from 1929 to 2001 (73 years) with 238 winners
and 528 nonwinning nominees. Sylvestre, Huszti and Hanley
(\citeyear{shh2006}) also used the survival analysis method suggested
by Efron (\citeyear{e2002}) and did an analysis with a binomial
logistic regression model. Death in each year of a performer's life was
treated as a Bernoulli random variable and regressed on covariates such
as winning status, age of nomination, and calendar year of nomination.
This model yielded a similar result as Sylvestre et al.'s Cox
proportional hazards model analysis. The results from previous studies
are listed in Table \ref{t1}.

\begin{table}
\caption{Winners vs. nominees}\label{t1}
\begin{tabular*}{\tablewidth}{@{\extracolsep{\fill}}lccc@{}}
\hline
 &   &  &\textbf{Reduction in}\\
&         &           &\textbf{mortality rate}\\
\textbf{Type of analysis}& \textbf{Status}        &\textbf{Time-zero}           &\textbf{(95\% CI) (\%)}\\ \hline
PH\tabnoteref{b1}           &Static\tabnoteref{b3}   &Birthday   &23 (3 to 39)\phantom{$-1$}\\
PH           &Dynamic\tabnoteref{b4}  &Birthday   &11 ($-12$ to 30)\\
PH           &Dynamic           &Nomination day  &18 ($-4$ to 35)\phantom{0}\\
PY\tabnoteref{b2}&Dynamic           &Nomination day  &18 ($-4$ to 36)\phantom{0}\\
\hline
\end{tabular*}
\legend{\textit{Notes}: These results are based on the updated
data set in Sylvestre, Huszti and Hanley
(\citeyear{shh2006}). The first row is the Cox model
without adjustment for any covariates; the second row is the Cox model
with winning status as a time-dependent covariate and with sex and year
of birth as time-independent covariates; the third row is the Cox model
with the same covariates as the second row, but with nomination day as
time-zero; the fourth row is the binomial logistic regression model
with sex, age, and calendar year as covariates. The first two rows are
from Redelmeier and Singh's analysis (using Sylvestre et al.'s updated
data set), and the last two rows are from Sylvestre et al.'s analysis.}
\tabnotetext[1]{b1}{PH stands for Cox proportional hazard model.}
\tabnotetext[2]{b2}{PY stands for performer years analysis, which is the binomial logistic
regression model described above.}
\tabnotetext[3]{b3}{Static status treats the winning status as a fixed-in-time
covariate.}
\tabnotetext[4]{b4}{Dynamic status treats the winning status as a time-dependent
covariate.}
\end{table}

\subsection{Healthy performer survivor bias}\label{s23}
Previous studies have suffered from healthy performer survivor bias,
that is, candidates who are healthier will be able to act in more films
and have more chances to win Oscar Awards.

One aspect of healthy performer survivor bias is \textit{immortal time
bias}, that is, candidates will have more chances to win Oscar Awards
if they live longer. When a performer is classified as a winner
throughout the study, regardless of when the performer wins the award,
there are unfair comparisons between winners and nonwinning performers
who died before the winner won the award. As an example, consider Henry
Fonda and Dan Dailey, who were both first nominated for an Oscar Award
at the age of 35 but did not win in their first nominations. Fonda
first won an Oscar at age 77 and died four months after, while Dailey
never won an Oscar and died at age 64. Fonda lived 13 years beyond the
age of Dailey's death before winning an Oscar. It is not fair to
consider the 13 years before Fonda won his Oscar as being affected by
winning.

To correct for immortal time bias, Sylvestre et al. used a Cox
proportional hazard model with the winning status as a time-dependent
covariate. In this model, the survival comparison between a winner and
a nonwinning nominee starts appropriately only at the time the winner
wins.

Although Sylvestre et al.'s analysis was an important advance in that
it corrects for immortal time bias, it still suffers from other aspects
of the healthy performer survivor bias. Winning an Oscar Award is an
indicator of being healthy. In Sylvestre et al.'s analysis, the risk
set at a given age consists of those performers who have been nominated
by that age. Among these performers, those who are healthy at the given
age have had more opportunities to perform and to win an Oscar. These
healthy performers are also more likely to live longer. Since having
won an Oscar is associated with survival in a risk set even if winning
has no causal effect on survival, there is the potential for bias.

As an example consider Jack Palance and Arthur O'Connell who were first
nominated for Oscars but did not win at ages 34 and 48, respectively.
Palance won an Oscar at age 73, while O'Connell never won an Oscar.
Palance was an active actor when he was in his 70s, acting in ten films
in his 70s, and lived to be 87. On the other hand, O'Connell was
stricken with Alzheimer's disease by the time he turned 70 and by the
time of his death at age 73, he was appearing solely in toothpaste
commercials (\href{http://www.imdb.com}{www.imdb.com}). The fact that Palance lived longer than
O'Connell in the risk set that started at age 73 after Palance's first
win is not likely due to the effect of winning but to the healthy
performer survivor bias.

One way of attempting to control for healthy performer survivor bias is
to condition on (control for) confounders in the Cox model. In
particular, nomination history is a confounder because it is a strong
risk factor for subsequently winning on Oscar Award (indeed, it is
necessary) and for mortality, since sick individuals do not get
nominated. Previous studies did not condition on nomination history and
thus suffered from confounding bias.

However, even if we condition on nomination history, as well as past
age and Oscar wins, and there are no other confounders besides these
variables, the time-dependent Cox model can be biased if Oscar winning
affects future nominations [Robins (\citeyear{r1986,r1992})]. It is substantively
plausible that previous Oscar winning affects future nomination (even
under the stronger null hypothesis that neither nomination nor winning
affects health). The effect could go in either direction. For example,
among two subjects with the same nomination history, only one of whom
won before, the winner would have a higher probability of being
renominated if increased fame coming from previously winning results in
an increased chance of nomination per film, all else being equal. On
the other hand, the winner would have a lower probability of being
renominated if nominators felt those who have not won before are more
deserving of a chance to win.

To understand the bias in the Cox analysis when previous Oscar winning
affects future nomination, suppose a previous winner has a higher
probability of being renominated, all else being equal. Then one would
expect that among the nominees in a given year with same past
nomination histories, the previous winners would be less healthy than
the previous nonwinners, since the nonwinners might have had to be in a
large number of movies in the previous year to get nominated for one of
them, while for the winner it often would suffice to be in just one.
But only a healthy person could be in many movies in one year. Note
that this bias persists even if we had data on the number of movies
performed in each year and adjusted for this variable as well as
nomination.

\subsection{Simulation studies}\label{s24}

To illustrate the potential of previous studies of survival in Oscar
Award winning performers to suffer from healthy performer survivor
bias, we conducted a simulation study.

We first assigned a lifetime for each performer and a time at when the
performer became sick. Then for each year, we randomly pick nominees
from performers who are still alive and healthy, and randomly select
one of them as the winner. Hence, winning an award does not have any
effect on prolonging performers' lifetime, because lifetime is
predetermined before deciding who wins the awards. If a method shows an
effect of winning over repeated simulations from this setting, it is
biased.

For each year between 1830 and 1999, we simulated five performers being
born. Each performer was randomly selected to have one of the three age
patterns shown in Table \ref{t2}.

\begin{table}
\caption{Performers' age pattern}\label{t2}%
\vspace*{-2pt}
\begin{tabular}{@{}ccc@{}}
\hline
&\textbf{Sick age}&\textbf{Death age}\\
\hline
Group 1&60&70\\
Group 2&70&80\\
Group 3&80&90\\
\hline
\end{tabular}
\vspace*{-6pt}
\end{table}

\begin{table}[b]
\vspace*{-4pt}
\caption{Selection weight for age 60--69}\label{t3}%
\vspace*{-2pt}
\begin{tabular}{@{}ccc@{}}
\hline
&\textbf{Previous winner}&\textbf{Previous nonwinner}\\
\hline
Group 1&0&0\\
Group 2&8&1\\
Group 3&9&7\\
\hline
\end{tabular}
\end{table}

For each year from 1927 to 2004, we have one award and we select 5
nominees from those performers who are still alive and healthy. The
details are that we select two nominees from the age group 30--39, and
one from 70--79, selecting randomly among healthy performers in those
age groups. We also select two nominees from the age group 60--69, but
with different selection probabilities for healthy performers in this
age group. For age group 60--69, the selection weight for a healthy
candidate is in Table \ref{t3}.

In this sense, for age group 60--69, winning in the past increases the
chance to be selected as a nominee, and previous nonwinners tend to be
healthier than previous winners (i.e., in Group 3 rather than in Group
2). This corresponds to the fact that previous nonwinners might have
had to be in a large number of movies in the previous year to get
nominated for one of them, while for the previous winners it often
would suffice to be in just one film to get nominated. Consequently,
nominated previous winners tend to be less healthy than nominated
previous nonwinners, because nominated previous nonwinners tend to be
very healthy to be able to act in many films.

Nominees from different age groups have a different probability to be
selected as the winner, with older nominees having a better chance. The
winning probability also depends on the nomination history and winning
history. Let $1_{30}$, $1_{60}$, and $1_{70}$ be the indicators of
current nomination age group 30--39, 60--69, and 70--79, respectively. Let
$N_{30}$, $N_{60}$, $N_{70}$ be the number of previous nominations in
the age group 30--39, 60--69, and 70--79, respectively. Let $W_{30}$,
$W_{60}$, $W_{70}$ be the number of previous wins in\vadjust{\goodbreak} the age group
30--39, 60--69, and 70--79, respectively. The winning probability for each
nominee in a given year is calculated as
\begin{eqnarray*}
&&P(A_i=1|\underline{N_i},\underline{A_i})\\
&&\qquad=\exp\bigl(0.5*1_{30}^i+1_{60}^i+2*1_{70}^i
\\
&&\qquad\quad\hphantom{\exp\bigl(}{}+
0.5(N_{30}^i+N_{60}^i+N_{70}^i+W_{30}^i+W_{60}^i+W_{70}^i)\bigr)
\\
&&\qquad\quad{}\Bigl/
\sum_{j=1}^5\exp\bigl(0.5*1_{30}^j+1_{60}^j+2*1_{70}^j
\\
&&\qquad\quad\hphantom{/\sum_{j=1}^5\exp\bigl(}{}+
0.5(N_{30}^j+N_{60}^j+N_{70}^j+W_{30}^j+W_{60}^j+W_{70}^j)\bigr).
\end{eqnarray*}
We choose these coefficients to magnify the healthy performer
survivor bias.

In our simulation setting, death ages are determined before winning,
thus winning has no causal effect on lifetime. Therefore, for the null
hypothesis that there is no treatment effect of winning an Oscar Award
on an actor's survival, the $p$-values should be uniformly distributed
between 0 and 1, and the mean of $p$-values should be around 0.5. If the
mean of $p$-values from a~method is much smaller than 0.5, then the
method is biased.

The results from 1000 simulations are shown in Table \ref{t4} and histograms
of $p$-values can be found in Figure \ref{f3} of Section \ref{s35}.

\begin{table}
\caption{Simulation results}\label{t4}
  \begin{tabular}{@{}lccc@{}}
  \hline
   \textbf{Type of analysis}   &\textbf{Status}   &\textbf{Time-zero}   &\textbf{Mean of }$\bolds{p}$\textbf{-value}\\
   \hline
     PH               &Static   &Birthday    &0.03\\
     PH               &Dynamic  &Birthday    &0.12\\
     PH               &Dynamic  &Nomination day  &0.12\\
     PY               &Dynamic  &Nomination day  &0.04\\
  \hline
   \end{tabular}
\end{table}

Redelmeier and Singh's results were based on the first two
methods in Table \ref{t4}, and Sylvestre et al.'s results were based on the
last two methods in Table \ref{t4}. All of these four methods are biased.

In our simulation setting, past winning history affects future
nominations, and past nomination history also affects future winning.
The previous methods did not account for the nomination history in the
time-dependent Cox model. Next we will show that even if one correctly
models the effect of nomination history on the hazard of death, the
hazard model still provides biased estimates of the causal effect of
winning on survival.

To simplify the consideration of nomination history and winning
history, we restrict every candidate to be nominated at most twice and
win at most twice. Let $D_{70}$ and $D_{80}$ denote death at age 70 and
80, respectively. Let $S_{69}$ and $S_{79}$ denote survival at age 69
and 79, respectively. Let $N(30)$ and $N(60)$ denote the numbers of
nominations in the age group 30--39 and 60--69, respectively. Let $A(30)$~and~$A(60)$ denote the numbers of wins in the age group 30--39 and
60--69,
respectively. Based on 1000 Monte Carlo simulations, we obtained
estimated mortality hazard rates and corresponding $95\%$ confidence
intervals for this full model in Tables \ref{t5} and \ref{t6}.

\begin{table}[b]
\tablewidth=230pt
 \caption{Mortality rates for death at 70 conditional
on survival to 69 with nomination history and winning history when
nomination is affected by past winning history}\label{t5}
  \begin{tabular}{@{}lcccc@{}}
  \hline
        &  & & &\textbf{Mortality rates}\\
           $\bolds{N(30)}$ &   $\bolds{A(30)}$     &   $\bolds{N(60)}$    &   $\bolds{A(60)}$  &\textbf{(95\% CI)}\\
           \hline
       2     & 2     &0    &0    &0.355 (0.299, 0.411)\\
      2     &1    &0   &0    &0.349 (0.332, 0.366)\\
      2   &0   &0   &0   &0.327 (0.320, 0.334)\\
      1   &1   &0   &0   &0.508 (0.494, 0.523)\\
      1   &0   &0   &0   &0.407 (0.404, 0.410)\\
      0   &0   &0   &0   &0.380 (0.378, 0.381)\\
  \hline
   \end{tabular}
\end{table}

\begin{table}
\caption{Mortality rates for death at 80 conditional on survival to 79
with nomination history and winning history when nomination is affected
by past winning history}\label{t6}
  \begin{tabular*}{\textwidth}{@{\extracolsep{\fill}}lcccccc@{}}
  \hline
        &  & & &  &  &\textbf{Mortality rates}\\
        $\bolds{N(30)}$  &  $\bolds{A(30)}$  &  $\bolds{N(60)}$  & $\bolds{A(60)}$   & $\bolds{N(70)}$  & $\bolds{A(70)}$  &\textbf{(95\% CI)}\\ \hline
      2   & 2  &0   &0   &0  &0  &0.472 (0.401, 0.544)\\
      2   &1   &0   &0   &0  &0  &0.511 (0.489, 0.534)\\
      2   &0   &0   &0   &0  &0  &0.493 (0.484, 0.502)\\
      1   &1   &0   &0   &0  &0  &0.555 (0.533, 0.577)\\
      1   &0   &0   &0   &0  &0  &0.674 (0.669, 0.678)\\
      1   &1   &1   &1   &0  &0  &0.474 (0.437, 0.511)\\
      1   &1   &1   &0   &0  &0  &0.468 (0.444, 0.492)\\
      1   &0   &1   &1   &0  &0  &0.191 (0.177, 0.206)\\
      1   &0   &1   &0   &0  &0  &0.190 (0.184, 0.196)\\
      0   &0   &1   &1   &0  &0  &0.439 (0.418, 0.460)\\
      0   &0   &1   &0   &0  &0  &0.521 (0.514, 0.528)\\
      0   &0   &2   &2   &0  &0  &0.137 (0.119, 0.155)\\
      0   &0   &2   &1   &0  &0  &0.092 (0.087, 0.098)\\
      0   &0   &2   &0   &0  &0  &0.039 (0.036, 0.041)\\
      0   &0   &0   &0   &0  &0  &0.617 (0.615, 0.619)\\
  \hline
   \end{tabular*}
\end{table}

For a reduced model without winning history, the mortality rates just
adjusting the nomination history are shown in Table \ref{t7}. From the
above probabilities, we can see even though winning has no causal
effect on survival, winning history affects the hazard of death given
nomination history, for example, the hazard of dying at 80 for people
with one nomination during their 30s and no further nominations is much
higher for people who did not win an award (0.674) than for those who
won one award (0.555).

If we consider a discrete time hazard model, the mortality rate can be
modeled as follows:
\[
h=\frac{1}{1+\exp(-\sum_i\alpha_iZ_i)},
\]
where $h$ is the mortality rate, and $Z_i$ is the indicator function of
nomination and winning history in the full model, or the indicator
function of nomination history in the reduced model. Then we can
estimate the coefficients $\alpha_i$ based on the mortality rates
calculated above. With this discrete time hazard model, we can
calculate the log likelihood of the full model and the reduced model
for each simulation round. Because
\[
-2\bigl(\mathit{loglikelihood}(\mbox{Reduced Model})-\mathit{loglikelihood}(\mbox{Full
Model})\bigr)\stackrel{D}{\rightarrow}\chi_{12}^2
\]
when the reduced model is true, we can obtain approximate $p$-values for
the test of whether winning has an effect on mortality given nomination
history. If the mean of $p$-values is significantly different from 0.5,
then it shows that even if one has a correct model for the conditional
hazard of death given all the measured time-dependent confounding
factors, the model still provides a biased estimate of the effect of
winning on survival.

\begin{table}
\tablewidth=230pt
\caption{Mortality rates conditional on nomination history when
nomination is affected by past winning history}\label{t7}%
\vspace*{-3pt}
  \begin{tabular}{@{}lcccc@{}}
  \hline
        &  & &  &\textbf{Mortality rates}\\
          \textbf{Death age}       &   $\bolds{N(30)}$   &   $\bolds{N(60)}$    &   $\bolds{N(70)}$    &\textbf{(95\% CI)}\\ \hline
   70            &2     &0      &       &0.331 (0.324, 0.337)\\
   70            &1     &0      &       &0.413 (0.410, 0.416)\\
   70            &0     &0      &       &0.380 (0.378, 0.381)\\[3pt]
   80            &2     &0      &0      &0.495 (0.487, 0.503)\\
   80            &1     &0      &0      &0.668 (0.664, 0.672)\\
   80            &1     &1      &0      &0.227 (0.222, 0.232)\\
   80            &0     &1      &0      &0.513 (0.506, 0.519)\\
   80            &0     &2      &0      &0.059 (0.057, 0.062)\\
   80            &0     &0      &0      &0.617 (0.616, 0.619)\\
  \hline
  \end{tabular}
  \vspace*{-5pt}
\end{table}

\begin{figure}[b]

\includegraphics{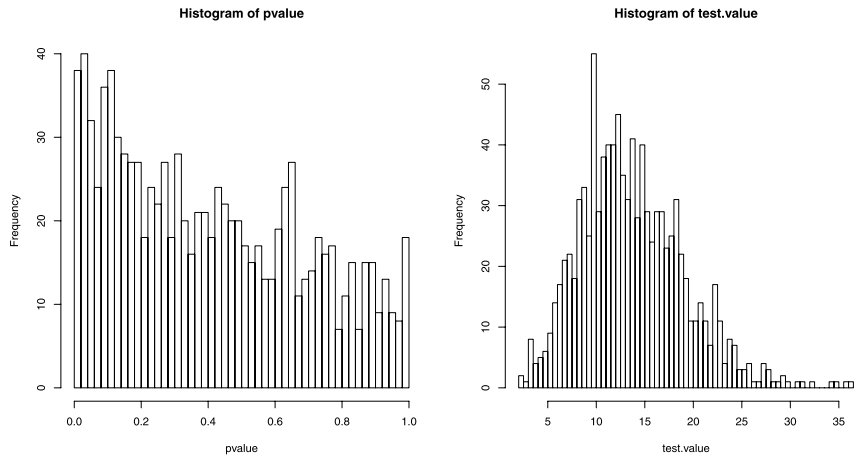}
\vspace*{-3pt}
\caption{Histograms for $p$-values and test statistics from the
likelihood ratio test of whether winning has an effect on mortality
given nomination history based on the discrete time hazard model when
the nomination is affected by past winning history.}\label{f1}
\vspace*{-5pt}
\end{figure}

The mean of $p$-values over 1000 simulation round is 0.404, showing that
there is bias. The histograms of $p$-values and test statistics are shown
in Figure \ref{f1}.

In the above simulation setting, nomination history is both a
confounder for winning history's effect on survival and has been
affected by winning history. We now show that if nomination history is
only a confounder and has not been affected by winning history, then
the time-dependent Cox model that controls for nomination history
produces correct inferences. We keep the same simulation set up as
before, except that we change the selection weights for age group 60--69
in Table~\ref{t3} to the selection weights in Table \ref{t8}.

\begin{table}
\tablewidth=194pt
\caption{Selection weight for age 60--69 when nomination history is not affected by winning history}\label{t8}
 \vspace*{-5pt}
  \begin{tabular}{@{}lcc@{}}
  \hline
  &\textbf{Previous winner}      &\textbf{Previous nonwinner}\\
  \hline
       Group 1     &0            &0\\
       Group 2     &8            &8\\
       Group 3     &9            &9\\
  \hline
   \end{tabular}
   \vspace*{-7pt}
\end{table}

We still restrict every candidate to be nominated at most twice and win
at most twice. Based on 1000 Monte Carlo simulations, we obtained
estimated mortality hazard rates for this full model in Table \ref{t9}.
\begin{table}[b]
\vspace*{-7pt}
\caption{Mortality rates conditional on nomination history and winning history when nomination is not affected by past winning history}\label{t9}
\vspace*{-5pt}
  \begin{tabular*}{\textwidth}{@{\extracolsep{\fill}}lccccccc@{}}
  \hline
   \textbf{Death}   & &  & & &  &  &\textbf{Mortality rates}\\
     \textbf{age}             &  $\bolds{N(30)}$  &    $\bolds{A(30)}$    &   $\bolds{N(60)}$    &  $\bolds{A(60)}$   & $\bolds{N(70)}$ & $\bolds{A(70)}$  &\textbf{(95\% CI)}\\ \hline
     70           &2   & 2  &0   &0   &   &     &0.317 (0.267, 0.368)\\
     70           &2   &1   &0   &0   &   &     &0.327 (0.310, 0.343)\\
     70           &2   &0   &0   &0   &   &     &0.337 (0.330, 0.344)\\
     70           &1   &1   &0   &0   &   &     &0.424 (0.411, 0.436)\\
     70           &1   &0   &0   &0   &   &     &0.425 (0.422, 0.429)\\
     70           &0   &0   &0   &0   &   &     &0.389 (0.388, 0.390)\\[3pt]
     80           &2   &2   &0   &0   &0  &0    &0.514 (0.449, 0.578)\\
     80           &2   &1   &0   &0   &0  &0    &0.508 (0.486, 0.529)\\
     80           &2   &0   &0   &0   &0  &0    &0.494 (0.485, 0.503)\\
     80           &1   &1   &0   &0   &0  &0    &0.592 (0.574, 0.611)\\
     80           &1   &0   &0   &0   &0  &0    &0.575 (0.570, 0.580)\\
     80           &1   &1   &1   &1   &0  &0    &0.461 (0.418, 0.504)\\
     80           &1   &1   &1   &0   &0  &0    &0.483 (0.454, 0.513)\\
     80           &1   &0   &1   &1   &0  &0    &0.481 (0.464, 0.498)\\
     80           &1   &0   &1   &0   &0  &0    &0.480 (0.472, 0.487)\\
     80           &0   &0   &1   &1   &0  &0    &0.686 (0.674, 0.697)\\
     80           &0   &0   &1   &0   &0  &0    &0.688 (0.683, 0.693)\\
     80           &0   &0   &2   &2   &0  &0    &0.428 (0.396, 0.459)\\
     80           &0   &0   &2   &1   &0  &0    &0.453 (0.440, 0.465)\\
     80           &0   &0   &2   &0   &0  &0    &0.451 (0.444, 0.458)\\
     80           &0   &0   &0   &0   &0  &0    &0.559 (0.558, 0.561)\\
  \hline
   \end{tabular*}
\end{table}
For a reduced model without winning history, the mortality rates just
adjusting the nomination history are shown in Table~\ref{t10}.
\begin{table}
\tablewidth=231pt
\caption{Mortality rates conditional on nomination history when nomination is not affected by past winning history}\label{t10}
\vspace*{-5pt}
  \begin{tabular}{@{}lcccc@{}}
  \hline
        & && &\textbf{Mortality rates}\\
            \textbf{Death age}     &   $\bolds{N(30)}$    &   $\bolds{N(60)}$    &    $\bolds{N(70)}$    &\textbf{(95\% CI)}\\ \hline
   70            &2     &0     &       &0.334 (0.328, 0.340)\\
   70            &1     &0     &       &0.425 (0.422, 0.428)\\
   70            &0     &0     &       &0.389 (0.388, 0.390)\\[3pt]
   80            &2     &0     &0      &0.499 (0.491, 0.507)\\
   80            &1     &0     &0      &0.577 (0.572, 0.581)\\
   80            &1     &1     &0      &0.480 (0.474, 0.487)\\
   80            &0     &1     &0      &0.687 (0.682, 0.691)\\
   80            &0     &2     &0      &0.451 (0.445, 0.457)\\
   80            &0     &0     &0      &0.559 (0.558, 0.561)\\
  \hline
  \end{tabular}
  \vspace*{-6pt}
\end{table}
From the probabilities in Tables \ref{t9} and~\ref{t10}, conditioning on the same
nomination history, winning does not have a~significant effect on the
mortality rates.

\begin{figure}[b]
\vspace*{-4pt}
\includegraphics{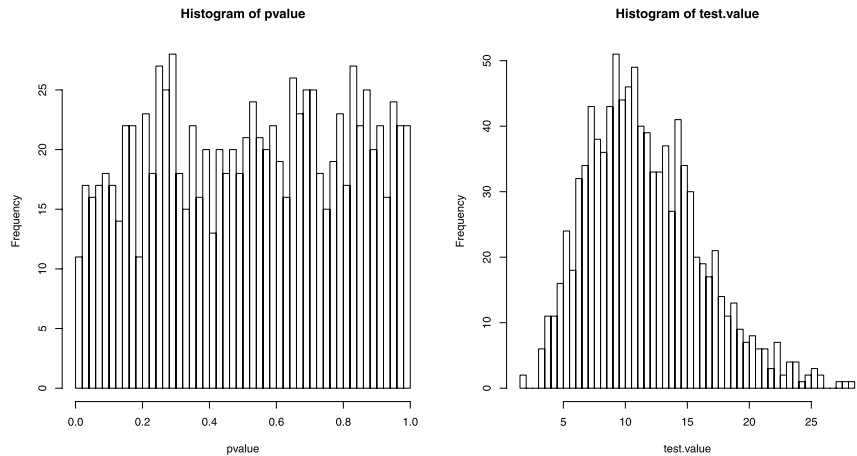}
\vspace*{-5pt}
\caption{Histograms for $p$-values and test statistics from the
likelihood ratio test of whether winning has an effect on mortality
given nomination history based on the discrete time hazard model when
the nomination is not affected by past winning history.}\label{f2}
\end{figure}

Similarly, based on the discrete time hazard model, the mean of
$p$-values in 1000 Monte Carlo simulations is 0.52, and the $p$-values and
test statistics of likelihood ratio test are shown in Figure \ref{f2}. The
simulation illustrates that when nomination is not affected by the past
winning history, a correct time-dependent hazard model does not suffer
from the healthy performer survivor bias.

\section{Rank preserving structural accelerated failure time
model}\label{s3}

Robins (\citeyear{r1986,r1992}) and Robins et al. (\citeyear{rbrw1992}) recognized the
potential of conventional time-dependent proportional hazard models to
provide biased estimates of causal effects when there are healthy
performer survivor effects (Robins called these healthy worker
effects). Robins (\citeyear{r1986}) was particularly concerned with occupational
mortality studies in which unhealthy workers who terminate employment
early are at an increased risk of death compared to other workers and
receive no further exposure to the chemical agent under study. More
generally, Robins has shown that the usual time-dependent Cox
proportional hazards model approach might be biased when ``(a) there
exists a time-dependent risk factor for, or predictor of, the event of
interest that also predicts subsequent treatment and (b) past treatment
history predicts subsequent risk factor level.'' In our context (a)
nomination history is a time-dependent risk factor for death and a
predictor of winning subsequent Oscar Awards, and (b) past winning
history predicts future nomination. Robins developed the rank
preserving structural accelerated failure time model with $g$-estimation
to eliminate bias from the time-dependent Cox proportional hazards
model under conditions (a) and (b) above. We will adapt Robins' rank
preserving structural accelerated failure time model and $g$-estimation
method.

Our key assumption is as follows:

\begin{asp}[(Randomization assumption)]\label{a1}
Conditional on age, previous nominations, and previous wins, the winner
of an Oscar Award in each year is selected randomly among nominees for
that award.
\end{asp}

We make no assumption about the nominees being randomly
selected from the pool of actors and actresses, only that the winner is
randomly chosen (conditional on covariates) among the nominees. Indeed,
some pundits suggest that being nominated for an Oscar Award is due to
talent, whereas winning one is due to luck [Sylvestre, Huszti and Hanley (\citeyear{shh2006})].
Gehrlein and Kher (\citeyear{gk2004}) provide further discussion of Oscar Award
selection procedures.

\subsection{Basic setup}\label{s31}

We focus on the causal effect of winning an Oscar Award for the first
time on a performer's survival, and do not consider any additional
effect of multiple wins here. We focus only on comparing winners to
nonwinning nominees.

To simplify our discussion, we use candidate $(i,j)$ to denote a
candidate $j$ who has been nominated for the $i$th Oscar Award. There
are a total of 300 Oscar Awards in our data, so $i=1,2,\ldots,300$. We
assume the existence of a \textit{latent or potential failure time
variable $U_{i,j}$}, which represents the potential years candidate
$(i,j)$ would live after the award date if he or she did not win an
Award on date $i$ nor in the rest of his or her lifetime. However, we
only observe the \textit{observed failure time variable $T_{i,j}$},
which means the observed years candidate $(i,j)$ lives after the award
date until his or her death. We will assume that the $T_{i,j}$ are
uncensored until Section \ref{s34}, where we will consider censoring.

\subsection{Rank preserving structural accelerated failure time
model}\label{s32}

The \textit{rank preserving structural accelerated failure time model}
(RPSAFTM) assumes that winning an Oscar for the first time multiplies a
performer's remaining lifetime by a treatment effect factor
$\exp(-\psi)$. The parameter $\psi$ is the additive effect of winning
on the log of a performer's remaining lifetime after the award. A
positive $\psi$ means winning decreases lifetime, a negative $\psi$
means winning increases lifetime and $\psi=0$ means winning has no
effect. See Cox and Oakes (\citeyear{co1984}) and Robins (\citeyear{r1992}) for more discussion
of the accelerated failure time model.

For the RPSAFTM, the potential failure time $U_{ij}$ can be calculated
from the observed failure time $T_{ij}$ as follows. Let $F_{i,j}$ be
the first time candidate $(i,j)$ won an Oscar Award ($F_{i,j}=\infty$
if the candidate never won an Award), and $D_i$ be the date of  the
$i$th Oscar Award. Let set $A$ contain candidates who never won an
Oscar Award in their whole lifetime, set $B$ contain candidates who won
Oscar Awards at least once and for whom $F_{i,j}<D_i$, and set $C$
contain candidates who won Oscar Awards at least once and for whom
$F_{i,j}\geq D_i$. We have
\begin{equation}\label{1}
\quad U_{i,j} =
\cases{
T_{i,j}, &\quad if candidate $(i,j)\in A\cup B$,\cr
F_{i,j}-D_i\cr
\qquad{}+\exp(\psi)(T_{i,j}+D_i-F_{i,j}), &\quad if candidate $(i,j)\in C$.}\hspace*{-8pt}
\end{equation}

As an example, consider Marlon Brando who was born on April 3, 1924,
and died on July 1, 2004. Brando was nominated for an Oscar for the
first time on March 20, 1952 ($i=77$), but did not win the Award. He
won two Oscar Awards in his career: the first time on March 30, 1955
($i=89$) and the second time on April 27, 1973 ($i=161$). His
information is listed in Table~\ref{t11}.
\begin{eqnarray*}
U_{77,\mathrm{B}}(\psi)&=&(30\mathrm{Mar}55-20\mathrm{Mar}52)+\exp(\psi)(1\mathrm{Jul}04-30\mathrm{Mar}55),
\\
U_{81,\mathrm{B}}(\psi)&=&(30\mathrm{Mar}55-19\mathrm{Mar}53)+\exp(\psi)(1\mathrm{Jul}04-30\mathrm{Mar}55),
\\
U_{85,\mathrm{B}}(\psi)&=&(30\mathrm{Mar}55-25\mathrm{Mar}54)+\exp(\psi)(1\mathrm{Jul}04-30\mathrm{Mar}55),
\\
U_{89,\mathrm{B}}(\psi)&=&\exp(\psi)(1\mathrm{Jul}04-30\mathrm{Mar}55),
\\
U_{101,\mathrm{B}}(\psi)&=&1\mathrm{Jul}04-26\mathrm{Mar}58,
\\
U_{161,\mathrm{B}}(\psi)&=&1\mathrm{Jul}04-27\mathrm{Apr}73,
\\
U_{165,\mathrm{B}}(\psi)&=&1\mathrm{Jul}04-2\mathrm{Apr}74,
\\
U_{231,\mathrm{B}}(\psi)&=&1\mathrm{Jul}04-26\mathrm{Mar}90.
\end{eqnarray*}

\begin{table}
\caption{Marlon Brando's nominations}\label{t11}
  \begin{tabular}{@{}lccc@{}}
  \hline
      \textbf{Nomination date }   &\textbf{Number of award} $\bolds{(i)}$    &\textbf{Award}    &\textbf{Win}\\
       \hline
       20Mar52    &\phantom{0}77      &Best Actor             &N\\
       19Mar53    &\phantom{0}81      &Best Actor             &N\\
       25Mar54    &\phantom{0}85      &Best Actor             &N\\
       30Mar55    &\phantom{0}89      &Best Actor             &Y\\
       26Mar58    &101                &Best Actor             &N\\
       27Apr73    &161                &Best Actor             &Y\\
        2Apr74    &165                &Best Actor             &N\\
       26Mar90    &231                &Best Supporting Actor  &N\\
  \hline
   \end{tabular}
\end{table}

The subscript ``B'' represents Marlon Brando. Note that in the RPSAFTM
(\ref{1}), Brando's multiple wins have no additional effect on his survival
beyond his first win.

\subsection{Test of treatment effect on survival}\label{s33}

Although the latent failure time variable $U_{i,j}$ can be calculated
based on the treatment effect factor $\psi$, $\psi$ is still an unknown
parameter that we need to estimate. The basic idea for testing the
plausibility of a hypothesized treatment effect under Assumption \ref{a1} is
the following: if the hypothesized treatment effect is correct, the
latent failure times in the treatment (winning) and control
(nonwinning) groups should be similar, but if the hypothesized
treatment effect is too large (small), the latent failure times in the
treatment group will tend to be smaller (larger) than those in the
control group.

To explain the details, let $A_{i,j}$ denote the treatment status for
candida\-te~$(i,j)$:
\[
A_{i,j} =
\cases{1, &\quad if candidate $(i,j)$ wins the $i$th award,\cr
0, &\quad if candidate $(i,j)$ loses the $i$th award.}
\]
Note that $A_{i,j}$ is only defined if $j$ was nominated for the $i$th
award. Let~$W_{i,j}$ denote the vector of candidate $(i,j)$'s
covariates, such as age at time of nomination, number of previous
nominations, and number of previous wins, etc. Note that some of the
covariates in $W_{ij}$ can be time dependent.

Let $U_{ij}(\psi_0)$ denote the latent failure time if $\psi_0$ is the
true treatment effect; $U_{ij}(\psi_0)$ can be calculated from (\ref{1}).
Consider a logistic regression model for the probability that candidate
$(i,j)$ wins award $i$ conditional on $W_{ij}$ and~$U_{ij}(\psi_0)$:
\begin{eqnarray}\label{2}
&&P\bigl(A_{ij}
=1|W_{ij},U_{ij}(\psi_0)\bigr)\nonumber
\\[-8pt]\\[-8pt]
&&\qquad=\frac{\exp(\beta
W_{ij}+\theta(\psi_0)U_{ij}(\psi_0))}{1+\exp(\beta
W_{ij}+\theta(\psi_0)U_{ij}(\psi_0))},\nonumber
\end{eqnarray}
where $\beta$ and $\theta(\psi_0)$ are unknown parameters. We use
conditional logistic regression for estimating (\ref{2}), where we
condition on there being one winner among the nominees for each award.
Only the nominees for each award are considered in the conditional
logistic regression, that is, the candidates included in the regression
are $(i,j_1),\ldots,(i,j_{n_i})$, where $i=1,\ldots,300$, and
$j_1,\ldots,j_{n_i}$ are the nominees for the $i$th award ($n_i=5$
except for some early awards). See the last two paragraphs of this
section for discussion of a~modification of this conditional logistic
regression that improves efficiency.\vadjust{\eject} Model~(\ref{2}) combined with
conditioning on there being one winner for each award is equivalent to
the model that the winner of award~$i$ is determined according to
McFadden's (\citeyear{m1974}) choice model where
$(W_{ij_1},\break U_{ij_1}(\psi_0)),\ldots,(W_{ij_{n_i}},U_{ij_{n_i}}(\psi_0))$
are the covariates that describe the $n_i$ choices for the award.

For the true $\psi$, the coefficient $\theta(\psi)$ on $U_{ij}(\psi)$
in (\ref{2}) should equal zero. This is because under Assumption \ref{a1},
conditional on the covariates $W_{ij}$'s of the nominees for an award,
the latent failure times $U_{ij}$'s of the nominees are independent of
which nominee wins the award, that is,
\[
P(A_{ij}=1|W_{ij},U_{ij})=P(A_{ij}=1|W_{ij}).
\]

We test the null hypothesis that $\psi$ equals a particular value
$\psi_0$ by seeing whether a score test accepts or rejects the null
hypothesis that the true value of $\theta(\psi)$ is 0. In other words,
we test
\[
H_{10}\dvtx \psi=\psi_0\quad \mathrm{vs.}\quad H_{\mathrm{1a}}\dvtx \psi\neq\psi_0
\]
by testing
\[
H_{20}\dvtx \theta(\psi_0)=0\quad \mathrm{vs.}\quad H_{\mathrm{2a}}\dvtx \theta(\psi_0)\neq0.
\]
Rejection of $H_{20}$ implies rejection of $H_{10}$, and acceptance of
$H_{20}$ implies acceptance of $H_{10}$. We invert this test to find a
confidence interval for $\psi$, that is, the $95\%$ confidence interval
consists of all $\psi_0$ for which we do not reject $H_{20}$.

We now discuss an efficiency issue for testing $\psi=\psi_0$. If a
candidate $(i,j)$ has already won an award before the date of the $i$th
Oscar Award, then $T_{ij}=U_{ij}$ regardless of whether the candidate
wins the award at the date of the $i$th Oscar Award. Candidate $(i,j)$
contributes no information for testing $\psi=\psi_0$ since
$U_{ij}(\psi_0)$ is a constant function of $\psi_0$. Consequently, it
is more efficient for testing $\psi=\psi_0$ to not include candidates
$(i,j)$ in the analysis who have already won an award before the date
of the $i$th Oscar Award. In fact, we found that for the Oscar data,
the confidence interval based on excluding candidates who have already
won an award was $20\%$ shorter than the confidence interval based on
including the already winners.

As an example of excluding the already winner candidates, for Marlon
Brando, we do not include $(101,\mathrm{B})$, $(161,\mathrm{B})$, $(165,\mathrm{B})$, $(231,\mathrm{B})$
because Brando won the $89$th Oscar Award (see Table \ref{t7}). Because we
estimate (\ref{2}) using conditional logistic regression in which we
condition on the number of winners for each award, by dropping
candidates $(i,j)$ who have already won an award before award $i$, we
effectively drop all data from awards in which the winner had already
won an award before.

\subsection{Censoring case}\label{s34}

If the lifetimes for all candidates were observed and Assumption \ref{a1}
holds, the above analysis would provide consistent tests for the
treatment effect. However, if some of the lifetimes are censored and we
treat the censored lifetime as the observed lifetime, there will be a
violation of Assumption \ref{a1}. Let $C_{i,j}$ denote the censoring time of
candidate $(i,j)$. For our data, $C_{i,j}={}$July 25, 2007 for all $i,j$.
Instead of observing the failure time~$T_{ij}$ of how long candidate
$j$ lives after the date $D_i$ of award $i$, we observe the censored
failure time $X_{ij}=\min(T_{ij},C_{ij}-D_i)$. Consider the variable~%
$U_{i,j}^*(\psi)$ that is generated by substituting $X_{i,j}$ for
$T_{i,j}$ in the RPSAFTM~(\ref{1}) to calculate $U_{i,j}$. If $\psi\neq0$,
then $U_{ij}^*(\psi)$ is not independent of $A_{ij}$ given~$W_{ij}$. To
illustrate this, we provide the following example. Suppose there is a
positive treatment effect for winning an Oscar Award on performers'
survival. Consider a candidate $A$ who just won once in his whole
career. Suppose he won on date $D$. Assume his actual remaining
lifetime after $D$ is $T$. If there is a positive treatment effect, his
latent failure time value will be $U$ where $U<T$. When the censoring
time $C$ satisfies $U<C-D<T$, the corresponding $U^*(\psi)$ generated
by substituting $C-D$ for $T$ in the RPSAFTM will be smaller than $U$
for the true $\psi$. Now consider a candidate $B$ who has the same latent
failure time $U$ and the same censoring time $C$ as candidate $A$, but
who never won any awards. For candidate $B$, we have $U^*(\psi)=U$.
Hence, for these two candidates with identical $U$'s, winning is
associated with $U^*(\psi)$. In summary, when there is a positive
treatment effect, winning an Oscar Award will prolong performers'
lifetime, making latent failure times more likely to get censored
compared to nonwinning nominees, and causing bias if censored failure
times are treated as actual failure times.\vadjust{\goodbreak}

In the above example, if we want to have the same censored latent
failure time for both winning and losing performers who have the same
actual latent failure time, we can modify the censoring time for the
losing performer to be before the actual censoring time so that
$U^*(\psi)$ will be censored in the same way regardless of whether a
performer wins or loses. This is Robins et al.'s (\citeyear{rbrw1992}) idea of
artificial censoring.

We define an observable variable $U_{i,j}^{**}(\psi_0)$ that is a
function of $(U_{i,j}(\psi_0),\break A_{i,j})$ and use it as a basis for
inference concerning $\psi_0$. $U_{ij}^{**}(\psi_0)$ is defined by
censoring $U_{ij}(\psi_0)$ at the artificial censoring time
$C_{ij}(\psi_0)$ that is defined below.

Recall that $F_{ij}$ is candidate $(i,j)$'s first win time, and $D_i$
is the date of~$i$th Oscar Award.

When $F_{ij}\geq D_i$,
\[
C_{i,j}(\psi_0)=\min\bigl((C_{i,j}-D_i),(C_{i,j}-D_i)\exp(\psi_0)\bigr).
\]

When $F_{ij}< D_i$,
\[
C_{i,j}(\psi_0)=C_{ij}-D_i.
\]

Then $U_{i,j}^{**}(\psi_0)=\min(U_{i,j}(\psi_0),C_{i,j}(\psi_0))$. We
substitute $U_{i,j}^{**}(\psi_0)$ for\break $U_{i,j}(\psi_0)$ in the
conditional logistic regression model (\ref{2}), and test the null
hypothesis $\theta(\psi_0)=0$. Note that $U_{ij}^{**}(\psi_0)$ could be
any observable function of $U_{ij}(\psi_0), C_{ij}(\psi_0)$, not just
$\min(U_{ij}(\psi_0),C_{ij}(\psi_0))$. Robins (\citeyear{r1993}) describes the
semiparametric efficient such function.

\subsection{Simulation results}\label{s35}

In Section \ref{s24} our simulation study showed that previous studies
suffered from healthy performer survivor bias. Here we will use the
same setup to test the \mbox{RPSAFTM}. Recall that a correct analysis method
should produce approximately uniformly distributed $p$-values in the
simulation study. The results in Table \ref{t12} are from 1000 simulations. We
have shown the first four rows from the simulations in Section \ref{s24}
(Table \ref{t4}), and add the last row for the \mbox{RPSAFTM}.\vadjust{\goodbreak}

\begin{figure}[b]
\vspace*{-5pt}
\includegraphics{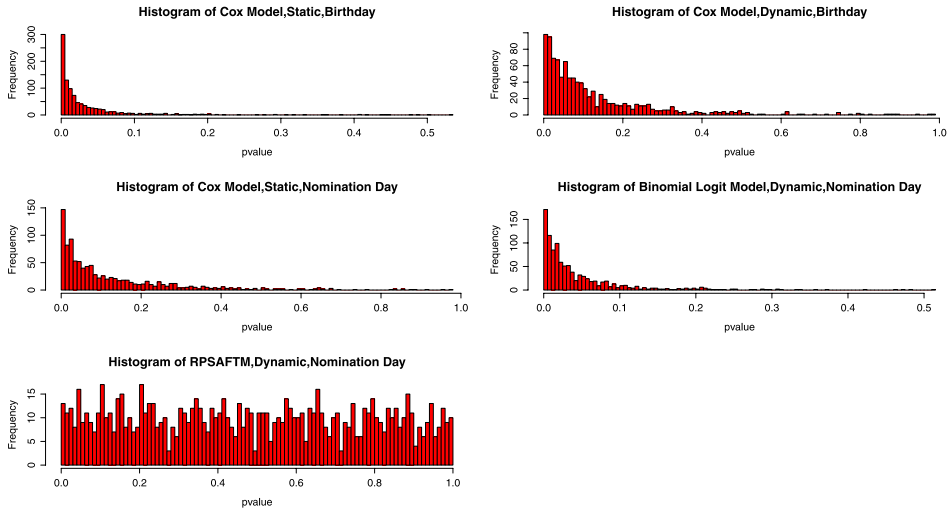}%
\vspace*{-5pt}
\caption{Histograms for $p$-values from the test of whether winning has
an effect on mortality based on the four methods of previous studies
and the RPSAFTM introduced by the current paper.}\label{f3}
\end{figure}

\begin{table}
\caption{Simulation results}\label{t12}%
\vspace*{-5pt}
  \begin{tabular}{@{}lccc@{}}
  \hline
   \textbf{Type of analysis}   &\textbf{Status}   &\textbf{Time-zero}   &\textbf{Mean of} $\bolds{p}$\textbf{-value}\\ \hline
     PH               &Static   &Birthday    &0.03\\
     PH               &Dynamic  &Birthday    &0.12\\
     PH               &Dynamic  &Nomination day  &0.12\\
     PY               &Dynamic  &Nomination day  &0.04\\
     RPSAFTM             &Dynamic  &Nomination day  &0.49\\
  \hline
   \end{tabular}
   \vspace*{-5pt}
\end{table}

Figure \ref{f3} contains histograms for $p$-values of the five methods from 1000
simulations.

In the first four plots, the majority of the $p$-values are smaller than
0.2, while in the last plot, the $p$-values are uniformly distributed.
The RPSAFTM corrects the survivor treatment selection bias that
previous methods suffer from.
\section{Analysis of Oscar Award data}\label{s4}

We have compiled a data file that records the nominees and winners for
each award (best lead actor, best lead actress, best supporting actor,
best supporting actress) on each Oscar Award date. We collected the
data from \href{http://www.imdb.com}{www.imdb.com}. The data is in the supplementary materials
[Han et al. (\citeyear{hsfp2010})]. The selection interval spanned
from the inception of the Oscar Awards to July 25, 2007. In computing
lifetime since being nominated, we use the actual Oscar Award date
which varies from year to year. People who were not reported dead on
\href{http://www.imdb.com}{www.imdb.com} were presumed to be alive. There are 260 winners and 564
nonwinning nominees, 824 performers in all. Of these 824 performers,
448 are censored.\looseness=1

We did not include several candidates in our data set. Margaret Avery
was nominated for best supporting actress in 1985, but we could not
find her birthday and day of death from the internet. We did not
include the following candidates who died before the winner of the
award for which they were nominated was announced: Massimo Troisi,
Jeanne Eagels, James Dean, Spencer Tracy, Peter Finch, and Ralph
Richardson.

We have shown results from previous studies, which are based on less
years of Oscar data than ours, in Table \ref{t1}. To compare previous studies
with ours, we have applied the methods of previous studies to our
updated Oscar Award data set; the results are shown in Table \ref{t13}.
Compared with the results in Table \ref{t1}, the reductions in mortality rate
in Table \ref{t13} are smaller. The confidence intervals are also narrower,
because we have 7 years more candidates than the original data set, and
also each candidate in our data set has 7 years more
information.

\begin{table}[b]
\caption{Winners vs. nominees}\label{t13}
  \begin{tabular}{@{}lccc@{}}
  \hline
       &  &  &\textbf{Reduction in}\\
                       &         &           &\textbf{mortality rate}\\
                    \textbf{Type of analysis}   &     \textbf{Status}     &     \textbf{Time-zero}      &\textbf{(95\% CI) (\%)}\\
                       \hline
          PH           &Static   &Birthday   &19 (6 to 31)\phantom{0.0..0$-$}\\
          PH           &Dynamic  &Birthday   &\phantom{0}9 ($-$6 to 22)\phantom{0.0..0}\\
          PH           &Dynamic  &Nomination day  &14 (0 to 26)\phantom{0.0..0$-$}\\
          PY           &Dynamic  &Nomination day  &10 ($-$6 to 23)\phantom{0.0..0}\\
          PH2           &Dynamic  &Nomination day  &\phantom{0}8.7 ($-$7.3 to 24.7)\\
   \hline
   \end{tabular}
\end{table}

In Table \ref{t8} the first four rows are based on previous methods. We also
add the fifth row, which corresponds to a Cox time-dependent model
adjusting for past nomination history and winning history; nomination
history is adjusted for by conditioning on the number of previous
nominations. Note that previous methods did not consider the nomination
history.

We now consider fitting the RPSAFTM. For the conditional logistic
regression (\ref{2}), we use the following time dependent covariates
$W_{ij}$: age of nomination (nomage), square of age of nomination
(nomage.square), cube of age of nomination (nomage.cubic), and number
of previous nominations (numprenom). Table \ref{t14} shows the results of the
conditional logistic regression model (\ref{2}) when $\psi=0$.

\begin{table}
\caption{Summary of conditional logistic model}\label{t14}
  \begin{tabular}{@{}lccccc@{}}
  \hline
              &\textbf{coef}    &$\mathbf{exp}$\textbf{(coef)}  &\textbf{se(coef)}  &$\bolds{z}$   &$\bolds{p}$\textbf{-value}\\ \hline
     $U_{ij}^{**}(0)$    &\phantom{$-$}$1.37e{-}$02  &1.01     &$0.007541$     &\phantom{$-$}1.812     &0.07\\
     nomage              &\phantom{$-$}$5.36e{-}$02     &1.06     &$0.101676$     &\phantom{$-$}0.527     &0.60\\
     nomage.square       &$-9.18e{-}$04    &1.00     &$0.002278$     &$-$0.403    &0.69\\
     nomage.cubic        &\phantom{$-$}$7.40e{-}$06     &1.00     &$0.000016$     &\phantom{$-$}0.462     &0.64\\
     numprenom           &\phantom{$-$}$6.99e{-}$02     &1.07     &$0.071407$     &\phantom{$-$}0.979     &0.33\\
  \hline
  \end{tabular}
\end{table}

The $p$-value for the test of whether the coefficient on $U_{ij}^{**}(0)$
is 0, that is, the test of $H_{20}\dvtx \theta(0)=0$ vs. $H_{2a}\dvtx \theta(0)\neq0$, is 0.07. Thus, we do \textit{not} reject the null
hypothesis that winning an Oscar has no effect on a performer's
survival at the 0.05 level. Looking at the effect of the other
covariates (the $W_{ij}$) in Table \ref{t14}, there is not strong evidence
that number of previous nominations has an effect on the probability of
a performer winning. For age at time of nomination, although the
$p$-values on each of the polynomial terms are not significant, a test
that the coefficient on all three of the terms is zero gives a~$p$-value
of 0.03 so age at time of nomination does appear to affect winning.
Older nominees are slightly more likely to win.

The validity of our test of the effect of winning an Oscar depends
critically on correctly controlling for the effect of age at time of
nomination on winning since this age is clearly correlated with
$U_{ij}$ (older nominees generally live a shorter time after the award
date, so have smaller $U_{ij}$'s). To check that our results are robust
to different ways of controlling for age at time of nomination, we
replaced the cubic polynomial in nomage in Table \ref{t10} with a~cubic spline
of nomage with 1 to 4 knots placed at equally spaced quantiles. The
$p$-values for the test of $H_{20}\dvtx \theta(0)=0$ vs. $H_{2a}\dvtx \theta(0)\neq0$ ranged from 0.064 to 0.07 in these analyses. Thus, our
result that there is not evidence that winning has an effect on
survival at the 0.05 level is robust to how nomage is controlled for.
We will use the cubic polynomial for nomage in Table \ref{t14} in our
subsequent discussion.

Table \ref{t15} shows the $95\%$ confidence interval for the treatment effect.
Our $95\%$ confidence interval is that the effect of winning is in the
range of decreasing survival (after the award date) by $0.88\%$ to
increasing survival by $26.62\%$.

\begin{table}
\tablewidth=185pt
\caption{$95\%$ confidence interval for treatment effect}\label{t15}%
\vspace*{-5pt}
  \begin{tabular*}{\tablewidth}{@{\extracolsep{\fill}}lc@{}}
  \hline
     \textbf{ Treatment effect}       &\textbf{CI}     \\ \hline
          $\psi$             &$[-0.2360,0.0088]$\\
      Winning multiplies survival  &  \\
      $\exp(-\psi)$        &$[0.9912,1.2662]$\\
  \hline
  \end{tabular*}
\end{table}

Robins' $g$-estimate for the treatment effect is the $\psi_0$ that makes
$\widehat{\theta}(\psi_0)=0$ in the conditional logistic regression
(\ref{2}). This $\psi_0$ maximizes the $p$-value for testing
$H_{20}\dvtx \theta(\psi_0)=0$ vs. $H_{2a}\dvtx \theta(\psi_0)\neq0$. Robins et
al. (\citeyear{rbrw1992}) show that the $g$-estimate is asymptotically normal and
consistent. The $g$-estimate can also be viewed as the Hodges--Lehmann
(\citeyear{hl1963}) estimate of the treatment effect based on the test of $H_{20}\dvtx \theta(\psi_0)=0$.

\begin{figure}[b]

\includegraphics{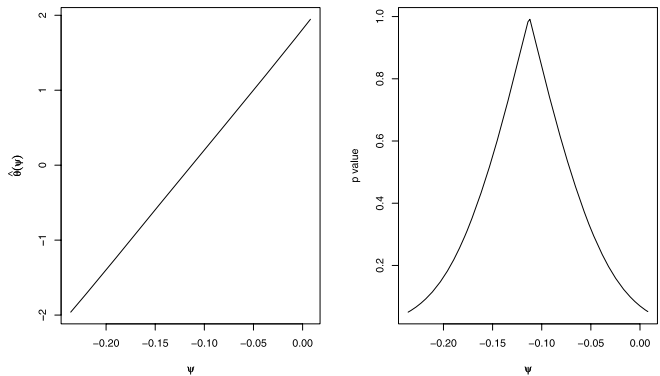}
\vspace*{-5pt}
\caption{Estimate of the coefficient $\theta$ of the modified potential
failure time variable $U_{ij}^{**}(\psi)$ in the conditional logistic
regression (\protect\ref{2}) for different treatment effect value $\psi$ and $p$-values
from the test of whether $\theta(\psi)$ equal zero for different
$\psi$.}\label{f4}
\end{figure}

We search for possible values of $\psi_0$ with
$\widehat{\theta}(\psi_0)=0$ in the range $[-0.2360,\break0.0088]$ with step\vspace*{1pt}
size${}=0.0001$. Figure \ref{f4} shows the estimates $\widehat{\theta}(\psi_0)$
and the $p$-values for testing $H_{20}\dvtx \theta(\psi_0)=0$.
$\widehat{\theta}(\psi)$ is a monotone increasing function of $\psi$ in
$[-0.2360,0.0088]$. The $g$-estimate is $\widehat{\psi}=-0.1127$, which
corresponds to winning increasing survival by $12\%$.
To estimate the survival advantage for winners in terms of years, we
consider the performers who won the first time they were nominated. For
these performers, we find their censored latent failure time
$U_{ij}^{**}(\widehat{\psi})$ under the assumption that the point
estimate $\widehat{\psi}$ of $\psi$ is the true treatment effect. Then
we make Kaplan--Meier estimates for the distribution of the actual
survival times for these winners and for the distribution of the latent
survival times if these winners had never won. The difference between
the estimated medians of these two distributions is an estimate of the
survival advantage of winning the award for these winners. In the
current Oscar Award data, we estimate the survival advantage to be 4.2
years, with a $95\%$ confidence interval of $[-0.4,8.4]$
years.

\subsection{Diagnostic plots}\label{s41}

To examine whether the RPSAFTM is appropriate for the Oscar Award data
set, we use boxplots to check if the randomization assumption
(Assumption \ref{a1}) is violated for latent failure times computed according
to the RPSAFTM at our point estimate $\widehat{\psi}$ of $\psi$. This
is similar to the diagnostics for testing an additive treatment effect
model in Small et al. (\citeyear{sgkr2006}). Based on the
randomization assumption, for the point estimate~$\widehat{\psi}$, the
distributions of $U_{ij}^{**}(\widehat{\psi})$ should be approximately
the same for the treatment group (winners) and the control group
(nonwinning nominees) in the same range of nomage. We divide the
candidates into five subgroups based on the quantiles of nomage. For
each subgroup, we make boxplots for $U_{ij}^{**}(\widehat{\psi})$ for
the winners and the nonwinning nominees.
\begin{figure}[b]

\includegraphics{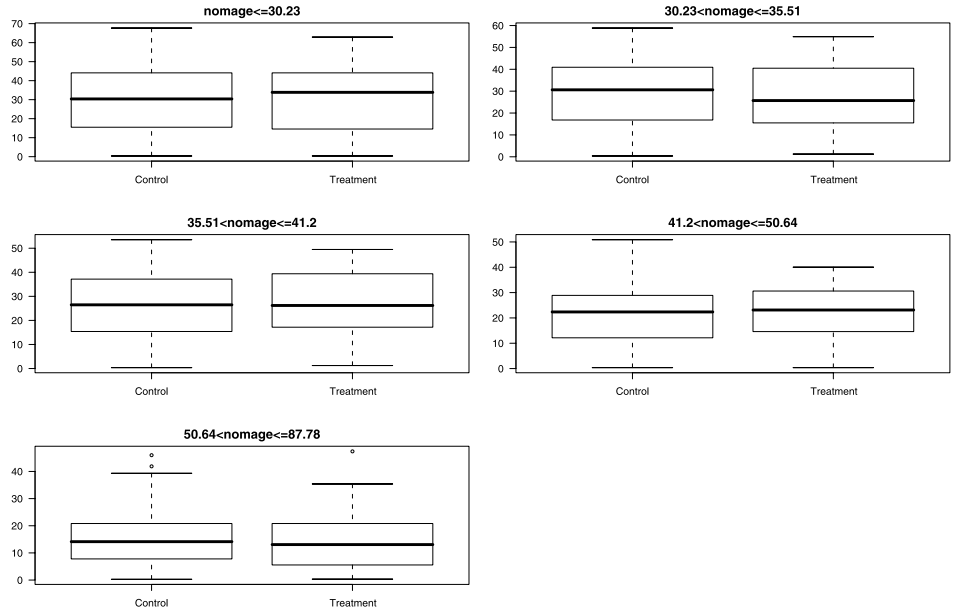}

\caption{Boxplots of $U_{ij}^{**}(\widehat{\psi})$ for comparison
between treatment group (winners) and control group (nonwinning
nominees) in five subgroups based on the quantiles of
nomage.}\label{f5}
\end{figure}
Figure \ref{f5} shows the distribution of $U_{ij}^{**}(\widehat{\psi})$ is
similar among winners and nonwinning nominees for each range of nomage.
This supports the validity of the RPSAFTM (assuming that Assumption \ref{a1}
is valid).

\subsection{Sensitivity analysis}\label{s42}

Our basic assumption, Assumption \ref{a1}, is that, conditional on covariates
such as age at nomination, and number of previous nominations, who wins
the Oscar Award is not related to how long the candidates would have
lived without winning an award. This could be violated if performers
who lead a more healthy lifestyle are more likely to win or if
performers who lead a more reckless lifestyle are more likely to win.
We now provide a sensitivity analysis to violations of Assumption \ref{a1}.
Under Assumption \ref{a1}, $\theta(\psi)$ is 0.\vadjust{\goodbreak} If Assumption~\ref{a1} is violated,
then $\theta(\psi)=\theta^{*}\neq0$. For $\theta(\psi)=\theta^*$, we
can test the plausibility of $\psi_0$ by testing $H_{20}'\dvtx \theta(\psi_0)=\theta^{*}$ vs. $H_{2a}'\dvtx \theta(\psi_0)\neq\theta^{*}$.
To calibrate $\theta^{*}$, we note that we can interpret
$\exp(10\theta^{*})$ as the odds ratio for one candidate to win
compared to another, if the one candidate has a ten year higher latent
failure time than the other and the two candidates are the same age at
nomination and have the same number of previous nominations. Under
Assumption \ref{a1}, $\exp(10\theta^*)=1$. Table \ref{t16} shows confidence intervals
for $\psi$ and the survival advantage of winning for winners at first
nomination for different values of $\theta^{*}$.

\begin{table}
\caption{Sensitivity analysis}\label{t16}
  \begin{tabular*}{\textwidth}{@{\extracolsep{\fill}}lccd{3.14}@{\hspace*{-2pt}}}
  \hline
      $\mathbf{exp}\!\bolds{(10\theta^*)=}$  & &  &\\
     \textbf{ Odds ratio for two}   &   &   &\multicolumn{1}{c@{}}{\textbf{Survival advantage}}\\
      \textbf{otherwise equal people}   &   &   &\multicolumn{1}{c@{}}{\textbf{in terms of years}}\\
      \textbf{one has 10 years  }       &  &  &\multicolumn{1}{c@{}}{\textbf{point estimate/}}\\
      \textbf{higher} $\bolds{U_{ij}^{**}}$ \textbf{than other}  & $\bolds{\theta^*}$&     \textbf{Confidence interval for} $\bolds{\psi}$    &\multicolumn{1}{c@{}}{\textbf{confidence interval}}\\
      \hline
      0.5     & $-$0.0693   &$(-0.6587,-0.4174)$   &16.4/(13.6,19.3)\\
      0.6     &$-$0.0511    &$(-0.5652,-0.3235)$   &14.1/(11.0,17.2)\\
      0.7     &$-$0.0357    &$(-0.4769,-0.2374)$   &11.7/(8.4,15.1)\\
      0.8     &$-$0.0223    &$(-0.3911,-0.1550)$   &9.3/(5.7,12.9)\\
      0.9     &$-$0.0105    &$(-0.3100,-0.0730)$   &6.8/(2.8,10.6)\\
      1\phantom{0.}       &\phantom{$-$}0\phantom{.0000}          &$(-0.2360,0.0088)$    &4.2/(-0.4,8.4)\\
      1.1     &\phantom{$-$}0.0095     &$(-0.1654,0.0879)$    &1.4/(-3.7,6.1)\\
      1.2     &\phantom{$-$}0.0182     &$(-0.0940,0.1697)$    &-1.5/(-6.9,3.6)\\
      1.3     &\phantom{$-$}0.0262     &$(-0.0210,0.2515)$    &-4.4/(-10.9,0.8)\\
      1.4     &\phantom{$-$}0.0336     &$(0.0413,0.3359)$     &\multicolumn{1}{c@{}}{$-7/(-16.4,-1.3)$}\phantom{$.\,$}\\
      1.5     &\phantom{$-$}0.0405     &$(0.0985,0.4238)$     &-10.3/(-19.2,-4.2)\\
   \hline
  \end{tabular*}
\end{table}

As the odds ratio $\exp(10\theta^*)$ increases from 0.5 to 1.5, the
point estimate of the survival advantage decreases from 16.4 years to
$-$10.3 years. If less healthy candidates are moderately more likely to
win than healthy candidates, $\exp(10\theta^*)=0.9$, then the
confidence interval only contains negative~$\psi$, and there is strong
evidence that winning increases survival. But if more healthy
candidates are somewhat more likely to win than less healthy
candidates, $\exp(10\theta^*)=1.2$, then the confidence interval
contains predominantly positive $\psi$ and the point estimate is that
winning decreases survival.

\section{Discussion}\label{s5}

In this paper we point out that healthy performer survivor bias exists
in methods from previous studies of the effect of winning an Oscar on
survival.\vadjust{\goodbreak} We show that under Assumption \ref{a1} (among nominees, the winner
is randomly selected conditional on baseline covariates), Robins'
RPSAFTM eliminates healthy performer survivor bias. We estimated that
the effect of winning an Oscar Award on survival for winners at first
nomination is to increase survival by 4.2~years, but the $95\%$
confidence interval of $[-0.4,8.4]$ years contains negative effects.
Thus, our study indicates that there is \textit{not} strong evidence
that winning an Oscar increases life expectancy.

The analysis in this paper is a case study of how Robins' RPSAFTM can
provide an improvement over Cox proportional hazards models for
estimating the effect on survival of a sudden change in a person's
life, for example, becoming ill, starting a high risk behavior, or
starting a treatment. A key assumption (our Assumption~\ref{a1}) that is
needed to obtain inferences from the RPSAFTM is that, conditional on
covariates recorded up to a given time, the sudden change is
``randomly'' assigned. A feature of our application, unlike most other
applications of RPSAFTMs [e.g., Robins et al. (\citeyear{rbrw1992}); Hern\'{a}n
et al. (\citeyear{hcmcr2005})], is that we only assume the sudden change is randomly
assigned among a select subset of the people in the study rather than
all people in the study. In particular, we are only assuming that among
nominees in a given year, who are generally at least somewhat healthy
in the given year, the winner is randomly selected.  We are not
assuming that the winner is randomly selected from the pool of all
actors and actresses who have been nominated in a previous year or the
given year and are still alive. Some performers nominated in a previous
year might be too unhealthy to act even though they are still alive.
Similar consideration of comparability only among a selected subset can
be found in Joffe et al. (\citeyear{jhjkcvr1998}) and Robins
(\citeyear{r2008}).\looseness=-1

In the RPSAFTM, model (\ref{1}) is rank preserving, that is, the effect of
winning is the same for each subject. Robins et al. (\citeyear{rbrw1992}) and Lok et
al. (\citeyear{lgvr2004}) discussed an expanded class of SAFTMs, which does not need
the RHS of (\ref{1}) at the true $\psi$ to be equal to the actual
counterfactual failure time $U_{ij}$, rather it just needs that the RHS
and the $U_{ij}$ have the same distribution conditional on past
measured covariates sufficient to control confounding. This eliminates
the assumption of rank preservation without changing the method of
estimation of the population (i.e., distributional) interpretation of
$\psi$.

\section*{Acknowledgments}
We thank James Robins for many insightful suggestions. We also thank
James Hanley, Marshall Joffe, and Paul Rosenbaum for helpful discussion
and suggestions. We thank Donald Redelmeier for providing the data he
used in his analysis. We thank the Associate Editor and the Editor for
many valuable suggestions which helped us in improving the paper.\vspace*{3pt}

\begin{supplement}
\sname{Supplement A}
\stitle{Oscar Award data for actors and actresses\\}
\slink[doi,text={10.1214/ 10-AOAS424SUPPA}]{10.1214/10-AOAS424SUPPA}
\slink[url]{http://lib.stat.cmu.edu/aoas/424/supplement.dat}
\sdatatype{.dat}
\sdescription{We have compiled a data file that
records the nominees and winners for each award (best lead actor, best
lead actress, best supporting actor, best supporting actress) on each
Oscar Award date. We collected the data from \href{http://www.imdb.com}{www.imdb.com}. The
selection interval spanned from the inception of the Oscar Awards to
July 25, 2007.}
\end{supplement}

\begin{supplement}
\sname{Supplement B}
\stitle{R code for data analysis and simulation\\}
\slink[doi]{10.1214/10-AOAS424SUPPB}
\slink[url]{http://lib.stat.cmu.edu/aoas/424/supplement.zip}
\sdatatype{.zip}
\sdescription{We provide the R code for our data
analysis and simulation studies. File ``R code.txt'' is for
preprocessing the Oscar data and data analysis in Section \ref{s4}. File
``simulation 1.txt'' is for the simulation studies in Sections \ref{s24} and
\ref{s35}, especially for Tables \ref{t4}, \ref{t12}, and Figure \ref{f3}. File ``simulation
2.txt'' is for the simulation studies in Tables~\ref{t5}--\ref{t10} and Figures \ref{f1} and \ref{f2}.}
\end{supplement}

\printaddresses

\end{document}